\documentclass[a4,12pt]{article}
\usepackage{graphicx,psfrag,feynmp,amssymb,cite,multirow}
\renewcommand{\thefootnote}{\fnsymbol{footnote}}

\newcommand{\prepr}[1] {\begin{flushright}  {\bf #1} \end{flushright} \vskip 1.cm}
\newcommand{\titul}[1] {\boldmath \begin{center}{\Large {\bf #1 } } \end{center}
\vskip 0.8cm}

\newcommand{\autor}[1] {\begin{center}  {\bf \lineskip .3cm #1  }
                        \end{center} }

\newcommand{\lugar}[1] {\begin{center}  {\normalsize \bf \it #1   } \end{center}}
\newcounter{muni}

\begin{document}
\begin{fmffile}{feynmangraphs}
\hbadness=10000
\pagenumbering{arabic}
\begin{titlepage}

\prepr{hep-ph/0205176\\
\hspace{30mm} KIAS--P02033 \\
\hspace{30mm} KEK--TH--824 \\
\hspace{30mm} May 2002}

\begin{center}
\titul{\bf $R$ Parity violating enhancement of $B_u^+\to \ell^+\nu$ and 
$B_c^+\to \ell^+\nu$}

\autor{A.G.\ Akeroyd$^{\mbox{1}}$\footnote{akeroyd@kias.re.kr},
S.\ Recksiegel$^{\mbox{2}}$\footnote{stefan@post.kek.jp} }
\lugar{ $^{1}$ Korea Institute for Advanced Study,
207-43 Cheongryangri-dong,\\ Dongdaemun-gu,
Seoul 130-012, Republic of Korea}

\lugar{ $^{2}$ Theory Group, KEK, Tsukuba, Ibaraki 305-0801, Japan }

\end{center}

\vskip2.0cm

\begin{abstract}
\noindent{We study the decays $B^+_u\to \ell^+\nu$ and 
$B^+_c\to \ell^+\nu$ in the context of the Minimal Supersymmetric 
Standard Model (MSSM) with explicit $R$ parity violation. We
analyse the correlation between the two decays and show that 
branching ratios (BRs) for $B^+_c\to \ell^+\nu$ may be of order 
5\% (over 40\% in one case), without violating current bounds 
on $B^+_u\to \ell^+\nu$. Although $B_c$ mesons are inaccessible at the 
$e^+e^-$ $B$ factories, such large BRs for $B^+_c\to \ell^+\nu$ would possibly 
be within experimental observability at LEP and the Tevatron Run II, 
with much larger yields expected at the hadronic $B$ factories.
We also update some earlier bounds on products of $R$ parity violating
couplings in the light of new experimental results.}

\end{abstract}

\vskip1.0cm
%{\bf  PACS index : }
\end{titlepage}
\thispagestyle{empty}
\newpage

\pagestyle{plain}
\renewcommand{\thefootnote}{\arabic{footnote} }
\setcounter{footnote}{0}

\section{Introduction}

Theoretical studies of rare decays of $b$ quarks have attracted
increasing attention since the start of the physics programme at the 
$B$ factories at KEK and SLAC. Both $B$ factories
are performing admirably and in excess of 60 fb$^{-1}$
of data has been accumulated by each experiment. 
The much anticipated measurement of
$\sin2\phi_1$ has established CP violation in the $B$ system 
\cite{Abe:2001xe, Aubert:2001nu}, while more recently 
evidence for CP violation in $B_d\to \pi^+\pi^-$ has been reported
\cite{Abe:2002qq}.

Many rare decays will be observed for the first time
over the next few years, and measurements of their branching ratios
(BRs) will be sensitive probes of models beyond the Standard Model (SM). 
In this paper we are concerned with the so far unobserved decays 
$B^+_u\to \ell^+\nu$ and 
$B^+_c\to \ell^+\nu$, where $\ell^+=e^+,\mu^+,\tau^+$ (and charged conjugates).
In the SM these purely leptonic decays proceed via annihilation of the 
$B$ meson to a virtual $W$, and offer the possibility of measuring the 
decay constant $f_B$. However, in the SM the rates are helicity
suppressed, and thus the largest BRs are for $\ell^+=\tau^+$.
The first measurement of BR$(B^+_u\to \tau^+\nu_{\tau}$) is 
expected at {\sc Belle} and {\sc BaBar}, while the SM BR$(B^+_u\to \ell^+\nu)$ 
for $\ell^+=e^+,\mu^+$ is below the experimental sensitivity.
$B_c$ mesons cannot be produced at the $e^+e^-$ $B$ factories operating
at the $\Upsilon(4S)$, but their search is possible at hadronic machines
(Tevatron Run II, LHC-B, BTeV). Although 
BR$(B^+_c\to \ell^+\nu)$ is considerably larger than the 
corresponding BR for $B^+_u$, the suppressed cross--section for
$B_c$ production and the larger backgrounds at hadron colliders
render isolation of these these decays very challenging, assuming the
BRs of the SM.

In models beyond the SM there can be sizeable enhancements from
new particles, which may bring some of these decays within experimental
observability.  In particular charged Higgs bosons ($H^\pm$) 
\cite{Hou:1992sy,Du:1997pm} and SUSY particles with $R$ parity violating 
interactions \cite{Guetta:1997fw,Baek:1999ch}, can both contribute at the
tree--level. We study the prediction for BR$(B^+_u\to \ell^+\nu)$
and BR$(B^+_c\to \ell^+\nu)$ in the MSSM with lepton violating
trilinear couplings $\lambda_{ijk}$ and $\lambda'_{ijk}$. In particular,
we examine the correlation between these decays and show that
by allowing a second $\lambda'$ coupling to be non--zero, the correlation
observed by previous authors can be avoided.
The BR$(B^+_c\to \ell^+\nu)$ can be enhanced to the level of
5\% (over 40\% in one case) without violating the bounds on BR$(B^+_u\to \ell^+\nu)$.
BRs of this magnitude may be in reach at the Tevatron Run II and LEP,
with plentiful yields at LHC-B and BTeV.
 
Our work is organised as follows: In section 2 we give a brief 
introduction to the $B_c$ meson. In section 3 we introduce the 
annihilation decays $B^+_u,B^+_c\to \ell^+\nu$, while section 4
examines the additional contributions to the decays in the MSSM with 
$R$ Parity violation. Section 5 presents the numerical results 
for the BRs and section 6 contains our conclusions.

%%%%%%%%%%%%%%%%%%%%%%%%%%%%%%%%%%%%%%%%%%%%%%%%%%%%%%%%%%
\boldmath
\section{The $B_c$ meson}\unboldmath
The $B_c$ meson is the last SM meson to be discovered
\cite{Gershtein:1994jw,Gershtein:1997qy}. It is
unique in the sense that it is composed of two heavy quarks of different
flavour. It possesses no strong or electromagnetic annihilation
decay channels because it carries open flavour.
Its mass and lifetime have been estimated by 
non--relativistic potential models and perturbative QCD
\cite{Brambilla:2001fw}, which predict $M_{B_c}\approx 6.2$ GeV 
and $\tau_{B_c}\approx 0.5 {\rm ps}$. 
Due to its mass it cannot be produced at the 
$e^+e^-$ $B$ factories running at the $\Upsilon(4S)$. It can, however, 
be produced at LEP and the Tevatron, which both
searched in the channels $B_c\to J/\psi \ell\nu_\ell, J/\psi \pi$.
LEP reported a few candidate events but claimed no signal 
\cite{Ackerstaff:1998zf,Abreu:1996nz,Barate:1997kk}.
The Tevatron Run I produced the first experimental evidence for the
$B_c$ meson, with a sample of 20 $B_c\to J/\psi \ell\nu_\ell$ events
corresponding to a background fluctuation of $4.8\sigma$,
\cite{Abe:1998wi,Abe:1998fb}. The measured mass and lifetime
were $6.4 \pm 0.4 GeV$ and $0.46^{+0.18}_{-0.16}\pm 0.03 ps$, 
respectively, in agreement with the theoretical predictions.
This is the only experimental information on
the $B_c$ meson, and hence the SM predictions for its BRs 
have not been verified. Large deviations from the SM rates
are still permitted by experiment. 

Although measurements with a precision comparable to the
$B_u$ and $B_d$ systems are not expected in the near future,
if new physics could affect the $B_c$ rates significantly
without contradicting existing measurements in other systems, 
these effects could be tested at the upcoming hadronic
$B$ machines where the $B_c$ meson accounts for approximately 
$10^{-3}$ of the inclusive cross-section for all beauty hadrons.

Possible decay channels for the $B_c$ meson are:
\begin{itemize}
\item[{(i)}] $c$ quark spectator, with $\overline b$ quark weak decay:
e.g.\ $B_c\to J/\psi \ell\nu_\ell, J/\psi \pi$
\item[{(ii)}]  $\overline b$ quark spectator, with $c$ quark weak decay
\item[{(iii)}] Annihilation decays: $B_c\to \ell\nu_\ell,\overline qq'$
\end{itemize}

In the SM, (i) and (ii) are expected to amount to around $90\%$ of 
the width, with the remaining $10\%$ coming from the annihilation decays. 
The latter is a much larger fraction than in the case of $B_u$ decays.
From now on we will concentrate on leptonic annihilation decays.

\boldmath
\section{The decays $B^+_u\to \ell^+\nu_\ell$ and $B^+_c\to \ell^+\nu_\ell$} \unboldmath

In the SM, the purely leptonic decays of $B_u^+$ and $B_c^+$ proceed via 
annihilation to a $W$ boson in the $s$-channel. The decay rate is given by
(where $q=u$ or $c$):
\begin{equation}
\Gamma(B^+_q\to \ell^+\nu_\ell)={G_F^2 m_{B_q} m_l^2 f_{B_q}^2\over 8\pi}
|V_{qb}|^2 \left(1-{m_l^2\over m^2_{B_q}}\right)^2
\end{equation}
Due to helicity suppression, the rate is proportional to $m^2_l$
and one expects:
\begin{equation}
BR(B^+_q\to \tau^+\nu_{\tau}):BR(B^+_q\to \mu^+\nu_{\mu})
:BR(B^+_q\to e^+\nu_{e})=m^2_{\tau}:m^2_{\mu}:m^2_e
\end{equation}
These decays are relatively much more important for $B_c$
than $B_u$ due to the enhancement factor $|V_{cb}/V_{ub}|^2
(f_{B_c}/ f_{B_u})^2$.
Searches have already been carried out for some of these decays.
The current experimental limits and SM predictions are presented
in Table \ref{explimits}.
\begin{table}\begin{center}
\begin{tabular} {|c|c|c|c|c|} \hline
Decay & SM Prediction & CLEO   & {\sc Belle} & LEP / Tevatron \\ \hline
 $B_u^+\to e^+\nu_e$ & $9.2\times 10^{-12}$ & $\le 1.5\times 10^{-5}$
 \cite{Artuso:1995ar}& 
 $\le 4.7\times 10^{-6}$\cite{BELLE:2001} & $\otimes$ \\ \hline
 $B_u^+\to \mu^+\nu_{\mu}$ & $3.9\times 10^{-7}$ & 
$\le 2.1\times 10^{-5}$ \cite{Artuso:1995ar}
  & $\le 6.5\times 10^{-6}$ \cite{BELLE:2001} & $\otimes$ \\ \hline
  $B_u^+\to \tau^+\nu_{\tau}$ & $8.7\times 10^{-5}$ 
& $\le 8.4\times 10^{-4}$ \cite{Browder:2000qr} & $\otimes$ 
& $\le 5.7\times 10^{-4}$ \cite{Acciarri:1996bv} / $\otimes$  \\ \hline
 $B_c^+\to e^+\nu_e$ & $2.5\times 10^{-9}$ & $\times$  
& $\times$ & $\otimes$ \\ \hline
 $B_c^+\to \mu^+\nu_{\mu}$ & $1.1\times 10^{-4}$  
 & $\times$ & $\times$ & $\otimes$ \\ \hline
 $B_c^+\to \tau^+\nu_{\tau} $ & $2.6\times 10^{-2}$ 
& $\times$ & $\times$ & (see text) / $\otimes$  \\ \hline
\end{tabular}\end{center}
\caption{SM predictions and current experimental limits from various machines. \label{explimits}}
\end{table}
``$\times$'' signifies that a search cannot be carried out since the
$B_c$ meson is kinematically inaccessible at the $B$ factories
CLEO, {\sc Belle} and {\sc BaBar}. The symbol ``$\otimes$'' signifies that no 
search has yet been performed, although in principle one could
have been carried out. This particularly applies to
LEP and the Tevatron which can produce both $B_u$ and $B_c$ mesons 
but have not searched for the majority of the annihilation channels.
CLEO and the $B$ factories have already reported strong limits on 
$B^+_u\to \ell^+\nu$ decays for $\ell^+=e^+,\mu^+$. The best limit on
$B^+_u\to \tau^+\nu$ decays comes from the L3 collaboration.
In the search for $B^+_u\to \tau^+\nu$ at LEP 
\cite{Acciarri:1996bv,Abreu:1999xe,Barate:2000rc}
there is a sizeable contribution from $B^+_c\to \tau^+\nu$, 
which has the same experimental signature 
\cite{Mangano:1997md}. Thus the experimental limit constrains  
the BR of an admixture of $B_u$ and $B_c$, and the effective
limit on the process $B^+_c\to \tau^+\nu$ depends strongly on
the conversion probability $b\to B_c$ 
(0.02\%--0.1\%) \cite{Mangano:1997md}. For lower estimates
of the conversion probability, the present experimental limit
puts a bound of the order of the SM width on $B^+_c\to \tau^+\nu$. 
Therefore future, more precise bounds on $B^+_u\to \tau^+\nu$
from high energy machines and/or a better knowledge of the 
$b\to B_c$ conversion probability will start to put meaningful 
constraints on BR$(B^+_c\to \tau^+\nu)$.

Neither the LEP collaborations nor
the Tevatron Run I searched for the channels $B^+_u\to {e/\mu}^+\nu$
and no limits whatsoever exist for $B_c^+\to \ell^+\nu$. 
While neither LEP nor Tevatron Run I could be competitive with the
$B$ factories in setting bounds for $B^+_u\to e/\mu^+\nu$, attempts to
measure this quantity could actually pick up a signal from
$B^+_c\to e/\mu^+\nu$. 
In the next section we shall show that in SUSY models with 
$R$ parity violation, the decays $B^+_c\to \ell^+\nu$
may be enhanced sufficiently to bring them within experimental 
sensitivity of both LEP and Tevatron, thereby motivating
a search in these channels. More specifically, assuming an
optimistic $b\to B_c$ conversion probability of $\approx 0.1\%$,
a measurement that gives a limit of $10^{-4}$ on the 
$B^+_u\to e/\mu^+\nu$ channels may be able to detect
$B^+_c\to e/\mu^+\nu$ BRs of ${\cal O}(5\%)$.

\boldmath
\section{$B^+_u, B^+_c\to \ell^+\nu_\ell$ in SUSY models with $R$ parity violation}
\unboldmath

Due to the current meagre experimental information on the decays
of $B_c$ mesons, future measurements of their BRs may be a test of 
models beyond the SM. It is entirely
possible that its BRs are very different from the SM predictions.
The expected yield of $B_c$ mesons from Run II at the Tevatron is $10^6$,
increasing to $10^9$ at LHC-B and BTeV \cite{Anikeev:2001rk,Ball:2000ba}.
These machines will provide
the first opportunity to study the $B_c$ meson in detail.

New physics contributions to the decays have been studied by a few
authors. \cite{Hou:1992sy} considered the effects of $H^\pm$
on the decays $B^+_u\to \ell^+\nu_\ell$, while \cite{Du:1997pm} extended this
analysis to the case of $B^+_c\to \ell^+\nu_\ell$. In both cases the $H^\pm$ 
contribution modifies the SM prediction by a factor $R$ where:
\begin{equation} \label{Rdef}
R=1-\tan^2\beta(M_{B_q}/M_{H^\pm})^2
\end{equation}
%Hence the rates for  $B^+_u\to \ell^+\nu_\ell$ and $B^+_c\to \ell^+\nu_\ell$
%are correlated. 

Since the SM predictions for
$B^+_u \to e^+\nu_e/ \mu^+\nu_\mu$ are already very small 
(see Table \ref{explimits}), the $H^\pm$ contribution
cannot enhance them to current experimental observability.
On the other hand, the $H^\pm$ contribution can saturate the
current experimental bounds on $B_u^+\to \tau^+\nu_{\tau}$.
The  DELPHI search for  $B_u^+\to \tau^+\nu_{\tau}$ constrains
$\tan\beta/M_{H^\pm} \le 0.46GeV^{-1}$ 
($90\%$ c.l.) which strengthens to $\tan\beta/M_{H^\pm} 
\le 0.42GeV^{-1}$ if the $B_c$ contribution is included \cite{Abreu:1999xe}.
L3 (who neglects the  $B_c$ contribution) \cite{Acciarri:1996bv} 
obtains $\tan\beta/M_{H^\pm}\le 0.38GeV^{-1}$, 
which would become stronger if the $B_c$ contribution were 
included in their analysis. The $B$ factories with $>100$ fb$^{-1}$
will be sensitive to BRs of the order of the SM prediction and thus the first 
measurement of $B_u^+\to \tau^+\nu_{\tau}$ is expected, unless its BR is 
suppressed by some new physics contribution ($H^\pm$ or otherwise).

The MSSM with $R$ Parity violating trilinear couplings predicts extra
contributions to $B^+_q\to \ell^+\nu$ from tree level 
slepton or squark exchange. It has been shown that these contributions
can be sizeable \cite{Baek:1999ch}.
The $R$ parity violating superpotential is given by 
\begin{equation}
W_{R}={1\over 2}\lambda_{ijk}L_iL_jE^c_k+\lambda'_{ijk}L_iQ_jD^c_k
     +{1\over 2}\lambda''_{ijk}U^c_iD^c_jD^c_k
\end{equation}

Here we do not consider the bilinear term $\mu_iL_iH_2$ since $\mu_i$ are 
constrained to be very small by neutrino masses \cite{Hirsch:2000ef}. 
Their contribution to $b$ decays
will be heavily suppressed compared to that from the trilinear couplings.

The simplest approach to $R$ parity violating phenomenology is
to assume that a single $R$ parity violating coupling in the
{\sl weak basis} is dominant with all others negligibly small. It was shown 
that such an approach leads to several non-zero $R$ parity violating
couplings in the {\sl mass basis} \cite{Agashe:1995qm}:
\begin{equation} \label{mixing}
\bar\lambda'_{ijk}=\lambda'_{imn}U_{Ljm}D^*_{Rnk}
\end{equation}

Where $\lambda'_{imn}$ represents the weak basis couplings
and $\bar\lambda'_{ijk}$ the mass basis.
In the SM, the left handed up--type and down type quark mixing matrices
$U_L$ and $D_L$ form the Kobayashi--Maskawa matrix $V_{KM}$. 
The right handed down type quark mixing matrix $D_R$ is unobservable in the
SM, and therefore the generation mixing in the $R$ parity violating couplings 
(\ref{mixing}) is a new observable. Limits on $R$ parity violating couplings
depend strongly on assumptions on the absolute quark mixing \cite{Agashe:1995qm,Dreiner:1997uz}.

We follow a frequently used approach and assume all the quark mixing to lie
in the up type sector. Thus $U_{Lij}=V^{KM}_{ij}$ and the above equation 
simplifies to:

\begin{equation}
\bar\lambda'_{ijk}=\lambda'_{imn}V^{\rm KM}_{jm}\delta_{kn}.
\end{equation}
This results in 3 non-zero $\bar\lambda'_{ijk}$.

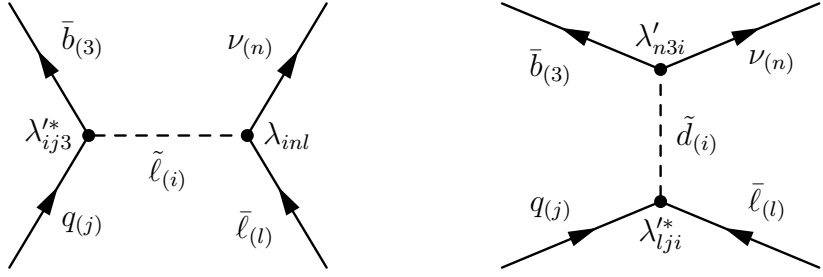
\begin{figure}
\begin{center}
\begin{fmfgraph*}(150,100)
\fmfleft{i1,i2} \fmfright{o1,o2}
\fmf{fermion,label=$q_{(j)}$}{i1,v1} 
\fmf{fermion,label=$\bar b_{(3)}$,label.side=right}{v1,i2} 
\fmf{fermion,label=$\bar \ell_{(l)}$,label.side=left}{o1,v2}
\fmf{fermion,label=$\nu_{(n)}$,label.side=left}{v2,o2}
\fmf{dashes,label=$\tilde \ell_{(i)}$}{v1,v2}
\fmflabel{$\lambda'^*_{ij3}$}{v1}
\fmflabel{$\lambda_{inl}$}{v2}
\fmfdot{v1,v2}
\end{fmfgraph*}
\hspace{1cm}
\begin{fmfgraph*}(150,100)
\fmfleft{i1,i2} \fmfright{o1,o2}
\fmf{fermion,label=$q_{(j)}$,label.side=left}{i1,v1} 
\fmf{fermion,label=$\bar b_{(3)}$,label.side=left}{v2,i2} 
\fmf{fermion,label=$\bar \ell_{(l)}$}{o1,v1}
\fmf{fermion,label=$\nu_{(n)}$}{v2,o2}
\fmf{dashes,label=$\tilde d_{(i)}$}{v1,v2}
\fmflabel{$\lambda'^*_{lji}$}{v1}
\fmflabel{$\lambda'_{n3i}$}{v2}
\fmfdot{v1,v2}
\end{fmfgraph*}
\end{center}
\caption{$R$ parity violating contributions to $B^+_q\to \ell_l^+\nu_n$. \label{feynmandiags}
Left: slepton exchange in the $s$--channel ($\sim\lambda_{inl}\lambda'^*_{ij3}$)
Right: squark exchange in the $t$--channel ($\sim\lambda'_{n3i}\lambda'^*_{lji}$).
Note that all four vertices conserve baryon number and violate lepton number,
this results in two fermion lines originating from (ending in) the upper
(lower) vertex of the right diagram.}
\end{figure}

In order to have $R$ parity violating contributions to the decays $B^+_q\to \ell^+\nu_\ell$ one must 
assume at least two non-zero trilinear terms in the weak basis. 
The possible contributions are shown in Figure \ref{feynmandiags}.

The $s$--channel contributions ($\sim\lambda\lambda'$)
are dominant and will from now on be considered exclusively. 
Unlike the $H^\pm$ case, there is no $m_l$ Yukawa coupling 
suppression for these contributions and thus large enhancements for 
BR($B_u^+/B_c^+\to l^+\nu$) for all lepton flavours are possible.

It was shown in \cite{Baek:1999ch} that the $\lambda\lambda'$ mediated 
contributions to the decay $B_q\to \ell\nu_\ell$ have the following couplings:
\begin{equation}
B^q_{ln}={\sqrt 2\over 4G_FV_{qb}}\sum_{i,j=1}^3{2\over 
m^2_{\tilde \ell_i}}V_{qj}\lambda_{inl}\lambda'^*_{ij3},
\end{equation} 
where $m_{\tilde \ell_i}$ is the mass of the slepton. 
For $q=u,c$ one has the explicit values:
\begin{eqnarray}
B^u_{ln} &=& \sum^3_{i=1}\lambda_{inl} \left(1689 \lambda'^*_{i13}
  + 384 \lambda'^*_{i23} + 1.52 \lambda'^*_{i33}\right) \nonumber \\
B^c_{ln} &=& \sum^3_{i=1}\lambda_{inl}  \left(32.7 \lambda'^*_{i13}
  + 144 \lambda'^*_{i23} + 6.1 \lambda'^*_{i33}\right)
\end{eqnarray}
(With $m_{\tilde \ell_i}=100 \,GeV$ and PDG--values for the CKM matrix elements.)

Using this approach, \cite{Baek:1999ch} showed that the upper 
limits on $B_u^+\to \ell^+\nu$ constrain several combinations of
$\lambda\lambda'$ each. They assumed a single non--zero $\lambda$ and
$\lambda'$ each in the weak basis. In this approach, the same couplings
mediate $B_u^+\to  \ell^+\nu$ and $B_c^+\to  \ell^+\nu$, and therefore the
experimental limits on $B_u^+\to \ell^+\nu$ induce limits on the
corresponding $B_c\to  \ell\nu$ channels which are not yet
experimentally constrained. Maximum values 
of BR$(B_c^+\to  \ell^+\nu_\ell)\approx 1\%$ were obtained. While
confirming the other results of \cite{Baek:1999ch}, in their approach we
obtain maximum values of BR$(B_c^+\to  \ell^+\nu)\approx 0.1\%$ only.
%\ref{XXX} {\bf FIX ME}
%This overestimation was due to a typographical error,
%and the authors of \cite{Baek:1999ch} have subsequently confirmed our
%substantially smaller results. 
% We do however, precisely reproduce their results when wrongly using
% $B^u_{ln}$ instead of $B^c_{ln}$ in the expression for BR$(B_c\to \ell\nu_\ell)$.

We will show that
in the minimal extension of this approach, i.e.\ 3 non--zero $R$ parity
couplings in the weak basis, this correlation can be dramatically
reduced, resulting in much larger BRs for $B_c\to \ell\nu_\ell\gg1\%$.

\section{Numerical results}

In Table \ref{couplingsandBRs} we give our results for the possible 
decay fractions
\begin{equation}
{\cal F}\equiv \Gamma(B^+_c\to\ell^+_l\nu_n)/\Gamma_{\rm tot}^{SM}.
\end{equation}
for the decays $B^+_c\to\ell^+_l\nu_n$. 
A value of ${\cal F}=100\%$ means that the BRs are 50\% for the combined SM channels
and 50\% for the $R$ parity violating contribution to the annihilation decay
in question; for small ${\cal F}$, ${\cal F}$ corresponds to a branching ratio.
We consider only the $R$--parity violating
contribution. The SM and charged Higgs $R$--parity 
conserving MSSM contributions (which can be up to 20 times as large
as the SM contribution) is negligible except for
$B^+_c\to\tau^+_l\nu_\tau$, which can reach up to 40\% for maximum
allowed $R$ (eq.\ \ref{Rdef}).

\begin{table}\begin{center}
\begin{tabular}{|c|c|c|c|c|c|c|}
\hline
Channel & Couplings used & $\lambda_{inl}$ & $\lambda'_{i13}$ & $\lambda'_{i23}$ & 
 ${\cal F}_{\rm opt}(\%)$ & ${\cal F}_{\rm con}(\%)$ \\
\hline
\multirow{6}{2.2cm}{$B^+_c\to e^+_l\nu_n$}
&$\lambda_{121} \lambda'_{1j3}$ & 0.05 & 0.02 & -0.088 & 13.77 & 0.004 \\
&$\lambda_{131} \lambda'_{1j3}$ & 0.06 & 0.02 & -0.088 & 19.82 & 3.305 \\
&$\lambda_{211} \lambda'_{2j3}$ & 0.05 & -0.041 & 0.18 & 57.53 & 0.003 \\
&$\lambda_{231} \lambda'_{2j3}$ & 0.06 & -0.041 & 0.18 & 82.85 & 3.305 \\
&$\lambda_{311} \lambda'_{3j3}$ & 0.06 & -0.045 & 0.20 & 102.2 & 0.003 \\
&$\lambda_{321} \lambda'_{3j3}$ & 0.06 & -0.045 & 0.20 & 102.2 & 0.004 \\
\hline
\multirow{6}{2.2cm}{$B^+_c\to \mu^+_l\nu_n$}
&$\lambda_{122} \lambda'_{1j3}$ & 0.05 & 0.02 & -0.088 & 13.76 & 0.001 \\
&$\lambda_{132} \lambda'_{1j3}$ & 0.06 & 0.02 & -0.088 & 19.81 & 4.953 \\
&$\lambda_{212} \lambda'_{2j3}$ & 0.05 & -0.041 & 0.18 & 57.50 & 0.004 \\
&$\lambda_{232} \lambda'_{2j3}$ & 0.06 & -0.041 & 0.18 & 82.80 & 4.953 \\
&$\lambda_{312} \lambda'_{3j3}$ & 0.06 & -0.045 & 0.20 & 102.2 & 0.004 \\
&$\lambda_{322} \lambda'_{3j3}$ & 0.06 & -0.045 & 0.20 & 102.2 & 0.001 \\
\hline
\multirow{6}{2.2cm}{$B^+_c\to \tau^+_l\nu_n$}
&$\lambda_{123} \lambda'_{1j3}$ & 0.05 & 0.02 & -0.088 & 11.65 & 4.193 \\
&$\lambda_{133} \lambda'_{1j3}$ & 0.004 & 0.02 & -0.088 & 0.075 & 0.075 \\
&$\lambda_{213} \lambda'_{2j3}$ & 0.05 & -0.041 & 0.18 & 48.68 & 2.796 \\
&$\lambda_{233} \lambda'_{2j3}$ & 0.06 & -0.041 & 0.18 & 70.09 & 70.09 \\
&$\lambda_{313} \lambda'_{3j3}$ & 0.004 & -0.045 & 0.20 & 0.385 & 0.385 \\
&$\lambda_{323} \lambda'_{3j3}$ & 0.06 & -0.045 & 0.20 & 86.54 & 4.193 \\
\hline
\end{tabular}\end{center}
\caption{Couplings and possible ${\cal F}\equiv \Gamma(B^+_c\to\ell^+_l\nu_n)/\Gamma_{\rm tot}^{SM}$. 
  \label{couplingsandBRs}
  ${\cal F}_{\rm opt}$: respecting single coupling bounds but ignoring the bounds from $B_d\to\ell^+\ell'^-$, 
  ${\cal F}_{\rm con}$: respecting the bounds from $B_d\to\ell^+\ell'^-$.
  PDG--values and $m_{\tilde \ell_i}=100 \,GeV$ have been used for the
  numerical parameters.}
\end{table}

The second to fourth columns give the values for the $R$--parity violating
couplings used in obtaining the optimistic
${\cal F}_{\rm opt}$. While
respecting single coupling bounds and the limits on $B^+_u\to\ell^+_l\nu_n$, 
some of the combinations $\lambda_{inl} \lambda'_{i13}$ used would violate the
bounds given by the experimental limits for $B_d\to\ell^+\ell'^-$. This could 
easily be avoided by allowing a fourth $R$--parity violating coupling to become
non--zero which can induce a cancellation analogous to the
$B^+_u\to\ell^+_l\nu_n$ case. We show the resulting 
${\cal F}_{\rm opt}$
in the fifth column of Table \ref{couplingsandBRs}.

Even when not demanding an additional cancellation and instead directly
respecting the bounds from $B_d\to\ell^+\ell'^-$, some of the combinations
can give quite respectable ${\cal F}$s (in one case up to 70\%) for 
$B^+_c\to\ell^+_l\nu_n$. These conservative ${\cal F}_{\rm con}$ 
are shown in the rightmost column. 
The larger surviving ${\cal F}$s of ${\cal O}(5\%)$ involve products of
$R$ parity violating couplings that are constrained by decays
involving one $\tau$ and one first or second generation lepton.
Since the experimental limits on these channels are typically
more than two orders of magnitude weaker than the ones involving
only first and second generation leptons, they imply less restrictive
bounds.

The largest ${\cal F}_{\rm con}$ in the rightmost column of Table \ref{couplingsandBRs}
($\lambda_{233} \lambda'_{2j3}$ inducing an 
${\cal F}_{\rm con}$ for $B^+_c\to\tau^+_l\nu_\tau$
of 70\%) is not constrained by purely leptonic decays of the $B_d$.
The potential bound for $\lambda_{233} \lambda'_{213}$ would come
from the decay $B_d\to\tau^+\tau^-$. For this decay no experimental
limits exist as yet,\footnote{{\sc BaBar} anticipates a sensitivity
at the level of $10^{-3}$ with a full dataset of $300fb^{-1}$.}
but it is estimated \cite{Grossman:1997qj,Guetta:1998fw}
that a limit of $1.5 \times 10^{-2}$ could be extracted from LEP
missing energy measurements. This limit would impose a bound of
$\lambda_{233} \lambda'_{213} < 3.0 \times 10^{-3}$, which is
smaller than the product of the single bounds $(5.4 \times 10^{-3})$,
but still does not further restrict our results involving this combination.

In finding ${\cal F}_{\rm con}$ we have used the results of \cite{Jang:1997jy}, updating
the bounds according to the new experimental limits for $B_d\to\ell^+\ell'^-$
\cite{BELLE:2001}.\footnote{While finalising this work, a paper appeared 
\cite{Saha:2002kt} that also attempts to update some of these bounds. 
The authors of that paper, however, do not use the latest $B$ factory limits 
\cite{BELLE:2001} which are considerably stricter than the ones that they employ.}
% Furthermore in many cases their ``previous bounds'' are less strict than even the
%``previous bounds'' (!) in JSL
%
The updated bounds are 
\begin{itemize}
\item $1.46 \times 10^{-5}$ instead of $4.6 \times 10^{-5}$
  for the combinations inducing $B_d\to e^+e^-$ \\ (i.e.\
  $\lambda_{121} \lambda'_{213}, \lambda_{121} \lambda'_{231}, 
  \lambda_{131} \lambda'_{313}, \lambda_{131} \lambda'_{331}$)
\item $1.79 \times 10^{-5}$ instead of $4.6 \times 10^{-5}$
  for the combinations inducing $B_d\to e^\pm\mu^\mp$ \\ (i.e.\
  $\lambda_{121} \lambda'_{113}, \lambda_{121} \lambda'_{131}, 
  \lambda_{121} \lambda'_{231}, \lambda_{122} \lambda'_{213}, 
  \lambda_{132} \lambda'_{313}, \lambda_{132} \lambda'_{331}, 
  \lambda_{231} \lambda'_{313}, \lambda_{231} \lambda'_{331}$)
\item $9.76 \times 10^{-6}$ instead of $2.4 \times 10^{-5}$
  for the combinations inducing $B_d\to \mu^+\mu^-$ \\ (i.e.\
  $\lambda_{122} \lambda'_{113}, \lambda_{122} \lambda'_{131}, 
  \lambda_{232} \lambda'_{313}, \lambda_{232} \lambda'_{331}$)
\end{itemize}
The other bounds are unchanged, because there are no new limits
for channels involving $\tau$ leptons or purely leptonic $B_s$ decays.

A more conservative approach that takes into account experimental uncertainties
in the input parameters is adopted in \cite{Dreiner:2001kc}. While using the
same CLEO limits on $B_d\to\ell^+\ell'^-$, they obtain weaker bounds on the
products of the $R$ parity violating couplings than \cite{Jang:1997jy,Baek:1999ch}. 
If we were to follow the approach of \cite{Dreiner:2001kc}, the above bounds
would weaken and the allowed ${\cal F}$s would become even larger.

For completeness we also update the bounds on combinations of
a single non--zero $\lambda$ and $\lambda'$ each as derived in \cite{Baek:1999ch}.
The new {\sc Belle} limits \cite{BELLE:2001} on $B^-_d\to e/\mu^-\nu$ change the bounds to
\begin{itemize}
\item $4.10 \times 10^{-5}$ instead of $7.3 \times 10^{-5}$
  for the combinations 
  $\lambda_{131} \lambda'_{113}, \lambda_{231} \lambda'_{213}$
\item $1.81 \times 10^{-4}$ instead of $3.2 \times 10^{-4}$
  for the combinations 
  $\lambda_{131} \lambda'_{123}, \lambda_{131} \lambda'_{323}, 
  \lambda_{231} \lambda'_{223}, \lambda_{231} \lambda'_{323}$
\item $1.14 \times 10^{-2}$ instead of $2.0 \times 10^{-2}$
  for the combination
  $\lambda_{231} \lambda'_{233}$
\item $4.82 \times 10^{-5}$ instead of $8.7 \times 10^{-5}$
  for the combinations 
  $\lambda_{132} \lambda'_{113}, \lambda_{232} \lambda'_{213}$
\item $2.12 \times 10^{-4}$ instead of $3.8 \times 10^{-4}$
  for the combinations 
  $\lambda_{132} \lambda'_{123}, \lambda_{132} \lambda'_{323}, 
  \lambda_{232} \lambda'_{223}$
\end{itemize}
Again, the the bounds on combinations that induce decays involving
$\tau$ leptons remain unchanged.

\begin{figure}
\begin{center}
\psfrag{113}{$\lambda'_{113}$} \psfrag{123}{$\lambda'_{123}$}
\psfrag{BR}{}\psfrag{optimistic}{\tiny choice for ${\cal F}_{\rm opt}$}
\psfrag{conservative}{\tiny choice for ${\cal F}_{\rm con}$}
\includegraphics[width=8cm]{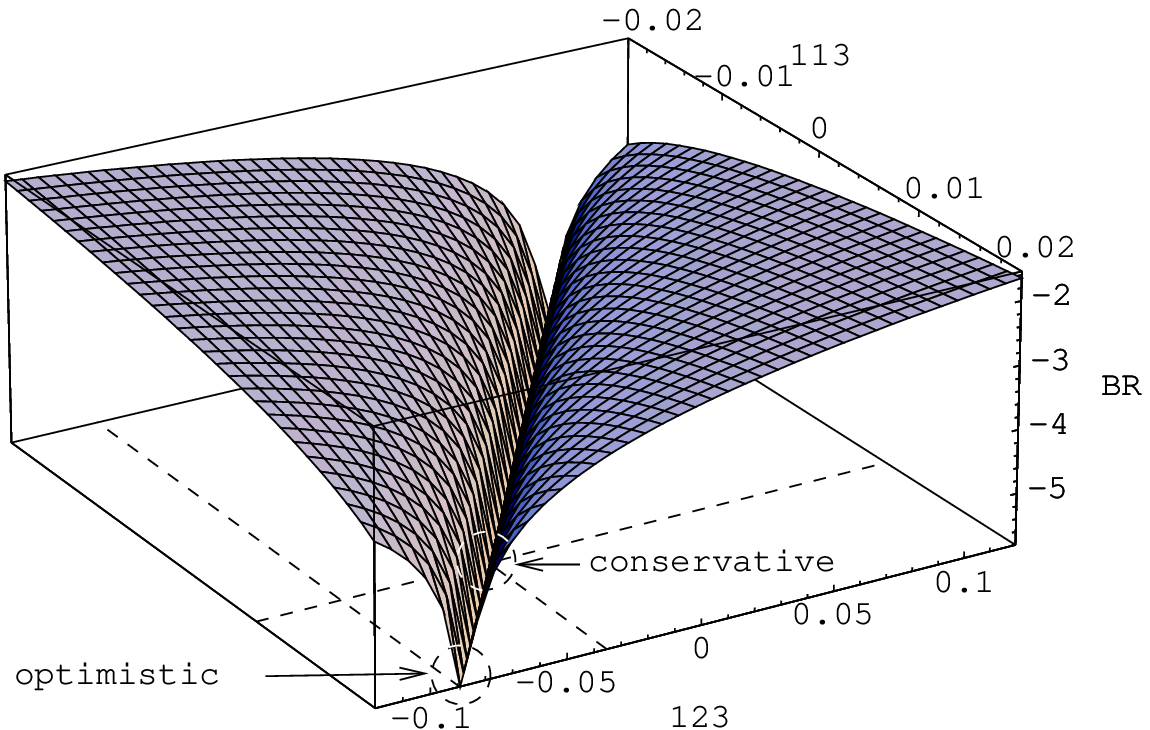}
\psfrag{123}{$\lambda'_{123}$}\psfrag{BR}{$\log_{10}BR(B^+_u\to\ell^+_l\nu_n)$}
\psfrag{ExpLimit}{\small Exp.\ limit} \psfrag{Interval}{\small Allowed interval}
\includegraphics[width=8cm]{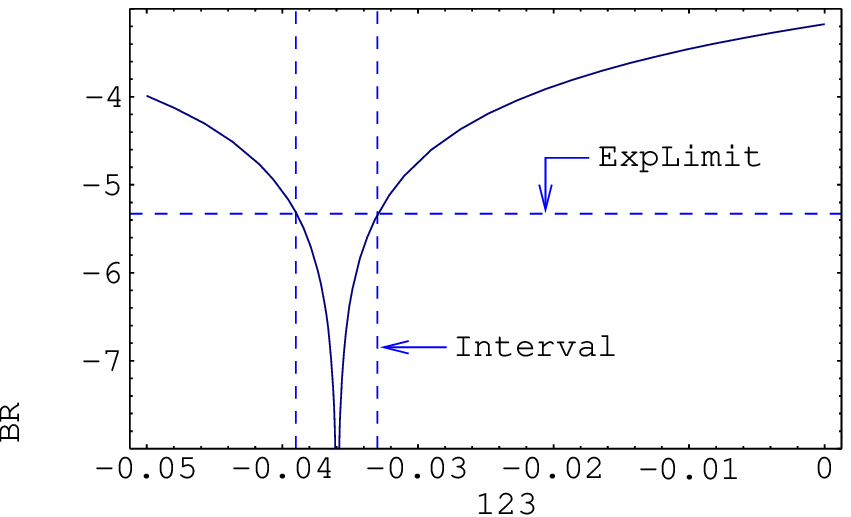}
\end{center}
\vspace*{-.5cm}
\caption{Cancellation in BR$(B^+_u\to\ell^+_l\nu_n)$. Left: BR against 
the $R$ parity violating couplings $\lambda'_{113}$ and $\lambda'_{123}$.
Right: BR against $\lambda'_{123}$ ($\lambda'_{113}=8.17\times 10^{-3}$),
indicating the experimental limit for BR$(B^+_u\to\ell^+_l\nu_n)$
and the allowed range for $\lambda'_{123}$. ($\lambda_{131} = 0.06$)}
\label{valleys}
\end{figure}

We would like to emphasise that the cancellation we employ to
respect the experimental bounds on $B^+_u\to\ell^+_l\nu_n$ is
not a very finely tuned one, e.g.\ let us look at the second line
of Table \ref{couplingsandBRs}. For the values quoted,
$\lambda_{131} = 0.06, \lambda'_{113} = 0.02$ and 
$\lambda'_{123} = -0.088$, the BR$(B^+_u\to e^+_l\nu_\tau)$
induced by this combination completely vanishes. In general,
there are areas in the  $\lambda'_{113}-\lambda'_{123}$ plane where
the individual contributions from these couplings interfere
destructively. We show this behaviour in the left part of Figure 
\ref{valleys}.

To respect the most recent {\sc Belle} bound, we have to
reduce BR$(B^+_u\to e^+_l\nu_\tau)$ to less than $4.7 \times 10^{-6}$.
To achieve this, $\lambda'_{123}$ could vary in the interval
$[-0.091,-0.085]$. By defining the amount of fine tuning
required for a certain cancellation as the width of the allowed 
interval over the distance of its middle from the origin,
this corresponds to a 7\% fine tuning. For the ${\cal F}$s in the rightmost 
column (which employ smaller $R$--parity violating couplings to respect
the $B_d\to\ell^+\ell'^-$ bounds), the fine tuning is even less severe.
For our example of the second 
line, $\lambda'_{123}$ can then vary in the interval $[-0.039,-0.033]$
which corresponds to a 17\% fine tuning.
The region in parameter space for which the desired cancellation
can occur, is therefore rather large.  We illustrate this
in the right hand part of Figure \ref{valleys}.
In general we find that the channels with
${\cal F}_{\rm con}\approx 5\%$ in Table \ref{couplingsandBRs} require a fine
tuning of $\approx 15\%$; for the ${\cal F}(B^+_c\to\tau^+\nu_\tau)=70\%$
we require a fine tuning of $4\%$.

%%%%%%%%%%%%%%%%%%%%%%%%%%%%%%%%%%%%%%%%%%%%%%%%%%%%%%%%%%

\section{Conclusions}
Purely leptonic decays of $B_d$, $B_u$, $B_s$ and $B_c$ mesons
are helicity (and in the neutral case also loop) suppressed 
in the SM with BRs typically smaller than
$10^{-10}$ for first generation leptons. Only
BR$(B^+_c\to\tau^+\nu_\tau)$ is expected to have a BR of ${\cal O}(3\%)$.
Combinations of two $R$ parity violating couplings can mediate
these decays without helicity suppression, and the experimental limits on
$B_{d/s}\to\ell^+\ell'^-$ and $B^+_u\to\ell^+_l\nu_{l'}$ induce
bounds on products of $R$ parity violating couplings that are
much stricter than the products of the individual bounds.
Since a single pair of non--zero couplings in the weak basis
can induce both $B^+_u\to\ell^+_l\nu_{l'}$ and
$B^+_c\to\ell^+_l\nu_{l'}$, these decays are correlated, and
the rather strict experimental bounds on $B^+_u\to\ell^+_l\nu_{l'}$
induce maximum BRs for the experimentally unconstrained decay channels
$B^+_c\to\ell^+_l\nu_{l'}$ of ${\cal O}(0.1\%)$ for the first
two generation leptons.

In this paper we have shown that in the minimal extension of
this two parameter approach, i.e.\ allowing three $R$ parity
violating couplings to be non--zero, the correlation between
the decays $B^+_u\to\ell^+_l\nu_{l'}$ and 
$B^+_c\to\ell^+_l\nu_{l'}$ is relaxed and the bounds on
the former decay do not restrict the latter anymore.
When directly respecting the bounds on products of two couplings
from $B_{d/s}\to\ell^+\ell'^-$ decays (``conservative approach''),
we obtain BRs of ${\cal O}(5\%)$ for channels involving first
and second generation leptons. Assuming an analogous cancellation
in the neutral $B$ meson leptonic decays by employing an additional
non--zero $R$ parity violating coupling (``optimistic approach''),
the partial width of these channels can even be of the order
of the total SM--width of the $B_c$ meson. For decays involving
$\tau$ leptons, the bounds from $B_{d/s}\to\ell^+\ell'^-$ are
so weak that even when respecting these bounds, ${\cal O}(1)$
BRs are possible.

\vspace{20mm}
\begin{center}
{\large\bf  Acknowledgements} 
\end{center}

The authors wish to thank F.\ Borzumati, Y.\ Okada and M.\ Yamaguchi
for useful discussions.
S.R.\ was supported by the Japan Society for the Promotion of Science (JSPS).

\renewcommand{\theequation}{B.\arabic{equation}}
\setcounter{equation}{0}

\end{fmffile}

\end{document}